\documentclass[twocolumn,superscriptaddress,letter,nobibnotes]{revtex4}
\usepackage{amssymb}
\usepackage{color}
\usepackage{graphicx}
\usepackage{dcolumn}
\usepackage{bm}
\usepackage[header,title,page,titletoc]{appendix}
\usepackage{amsmath}
\usepackage{amsfonts}

\makeatletter
\newcommand{\rmnum}[1]{\romannumeral #1}
\newcommand{\Rmnum}[1]{\expandafter\@slowromancap\romannumeral #1@}
\makeatother

\begin{document}

\title{Non-Hermitian Chiral Skin Effect}
\author{Xin-Ran Ma}
\thanks{Who has the same contribution to this work}
\affiliation{Center for Advanced Quantum Studies, Department of Physics,
Beijing Normal University, Beijing 100875, China}
\author{Kui Cao}
\thanks{Who has the same contribution to this work}
\affiliation{Center for Advanced Quantum Studies, Department of Physics,
Beijing Normal University, Beijing 100875, China}
\author{Xiao-Ran Wang}
\affiliation{College of Teacher Education, Hebei Normal University,
Shijiazhuang 050024, China}
\author{Zheng Wei}
\affiliation{Center for Advanced Quantum Studies, Department of Physics,
Beijing Normal University, Beijing 100875, China}
\author{Qian Du}
\affiliation{Center for Advanced Quantum Studies, Department of Physics,
Beijing Normal University, Beijing 100875, China}
\author{Su-Peng Kou}
\email{spkou@bnu.edu.cn}
\affiliation{Center for Advanced Quantum Studies, Department of Physics,
Beijing Normal University, Beijing 100875, China}

\begin{abstract}
The interplay between non-Hermitian effects and topological insulators has become a frontier of research in non-Hermitian physics. However, the existence of a non-Hermitian skin effect for topological-protected edge states remains controversial. In this paper, we discover an alternative form of the non-Hermitian skin effect called the non-Hermitian chiral skin effect (NHCSE). NHCSE is a non-Hermitian skin effect under periodic
boundary condition rather than open boundary condition. Specifically, the chiral modes of the NHCSE localize around \textquotedblleft topological defects\textquotedblright characterized by global dissipation rather than being confined to the system boundaries. We show its detailed physical properties by taking the non-Hermitian Haldane model as an example. As a result, the intrinsic mechanism of the hybrid skin-topological effect in Chern insulators is fully understood via NHCSE. Therefore, this progress will be helpful for solving the controversial topic of hybrid skin-topological effect and thus benefit the research on both non-Hermitian
physics and topological quantum states.
\end{abstract}

\maketitle

\section{Introduction}

The non-Hermitian quantum systems have attracted intensive attention due to
their effectiveness in describing non-equilibrium and open systems, as well as their rich underlying physics distinguishing them from Hermitian
counterparts\cite{Bender98, Shen2018, Liu2019, xi2019, QWG1, QWG2, NHPH,
GuoC2020, WWang2022, Bergholtz2021}. There are many intriguing phenomena in
non-Hermitian quantum systems, such as exceptional degeneracies\cite%
{Bender98, EP1}, unidirectional transmission\cite{trans_1, trans_2} and
non-Hermitian skin effect (NHSE) \cite{Yao2018, Yao20182, Xiong2018,
Torres2018, Ghatak2019, Lee2019, Kunst2018, Yokomizo2019, Yin2018,
KawabataUeda2018, SongWang2019, Longhi2019, KZhang2020, Slager2020, YYi,
Okuma2021, Roccati, Shen Mu, E.Lee, Tliu, bose mb1, bose mb2, CF2022,
HLiu2022, gong2019, zhangxd_1}. The NHSE has the eigenstates of the bulk
that are exponentially localized at the edge of the system with an open boundary condition (OBC) and can be characterized by the generalized
Brillouin zone (GBZ) rather than the usual Brillouin zone (BZ) \cite%
{Yao2018, Lee2019, Yokomizo2019, Yokomizo2022}. In addition, the NHSE is
pointed out to be relevant to point-gap topology and be identified by a
corresponding \textquotedblleft \emph{winding number}\textquotedblright \cite%
{KZhang2020, okuma2020}. The existence of the \textquotedblleft winding
number\textquotedblright guarantees the NHSE.

The investigation of the interplay between the NHSE and topological insulators has yielded a plethora of captivating physical phenomena. These include the emergence of defective edge states \cite{wangxr2020-1, wangxr2020-2}, the development of non-Hermitian topological invariants \cite{Kunst2018, Yao2018, Ghatak2019, KZhang2020,
lil2023spectral_winding}, as well as the manifestation of synergy and hybridization between the NHSE and band topology in higher dimensions \cite{xi2019, high_order_kawabata2020, HLiu2022, gong2019, X.Zhang_hybrid, gong_haldane, zhangxd_2, high_order_Nori2019,
li2023topological, zhu2023photonic, high_order_chris2019, high_order_Bergholtz2019}. Specifically, the hybrid skin-topological effect, which represents a distinct manifestation of NHSE, manifests exclusively in topological edge states and is confined to systems of two and higher dimensions. This effect results in the localization of topologically protected edge states at specific corners while the bulk states are extended. Previous investigations have found two types of the hybrid skin-topological effect: those induced by non-reciprocal lattice configurations \cite{gong2019, zhangxd_2, Li_2020_hy_exp} and those induced by gain and loss \cite{HLiu2022}. Notably, an important observation has been documented in a two-dimensional Chern insulator, where chiral edge states are localized at certain corners due to gain/loss \cite{HLiu2022}. This observation holds significant implications as chiral states exhibit robustness against backscattering and Anderson localization in Hermitian
systems. Nevertheless, a comprehensive understanding of the underlying
mechanisms governing the gain-loss-induced hybrid skin-topological effect
remains challenging.

In this paper, we try to solve this puzzle after answering the following
questions: (\romannumeral 1) Is there a fundamental distinction between the NHSE that occurs in bulk states and that appears in topological edge states of a Chern insulator? (\romannumeral 2) If such a distinction exists, how
can we determine whether NHSE occurs in topological edge states, and how to
characterize it? (\romannumeral 3) What is the physical basis for these
distinctions, and how do they relate to the hybrid skin-topological effect?
Can we have a deeper understanding of the hybrid skin-topological effect
with the help of NHSE for topological edge states?

To answer the above questions, we investigate the non-Hermitian chiral skin effect (NHCSE) under periodic boundary conditions (PBC) by studying chiral modes with inhomogeneous, perturbative dissipation. Specifically, we analyze the non-Hermitian Haldane model and examine the influence of global dissipation's domain walls (GDDWs) as a form of inhomogeneous dissipation. Our focus is on two key properties of NHSE: the anomalous occurrence exclusively of GDDWs in topological systems of two or higher dimensions, and the phenomenon of boundary reconstruction in cylindrical Chern insulators with GDDWs on one edge. Moreover, our study presents a novel mechanism for the hybrid skin-topological effect by introducing the concept of the NHCSE.

This work is organized as follows. Sec. \ref{sec.2} provides an introduction to the theory of the NHCSE. In Sec. \ref{sec.3}, we investigate the non-Hermitian Haldane model with inhomogeneous and perturbative dissipation. Specifically, we investigate the cases of bulk dissipation in Sec. \ref{case(a)} and edge dissipation in Sec. \ref{case(b)} to explore the manifestation of the NHCSE. In Sec. \ref{sec.6}, we use NHCSE to understand the hybrid skin-topological effect. Moreover, in Sec. \ref{sec.7}, we introduce a circuit design capable of implementing the NHCSE in the Haldane model with GDDWs. Finally, we conclude our work in Sec. \ref{sec.8}.

\section{Non-Hermitian chiral skin effect}
\label{sec.2}
Firstly, we show the key properties of the NHCSE for chiral modes with
inhomogeneous dissipation. We emphasize that NHCSE is a unique type of NHSE under PBC rather than OBC.

In the continuous limit, the effective single-body Hamiltonian for chiral
modes in low-energy physics becomes
\begin{equation}
h_{\mathrm{chiral}}=v\cdot k,
\end{equation}
where $v$ is the velocity of edge states, and $k$ is the wave vector of the
chiral modes. There are two types of chiral modes, one with positive
velocity $v>0,$ the other with negative velocity $v<0.$ Then, we consider
the effect of dissipation, of which the effective strength is $\gamma $ ($%
\text{Im}\gamma \equiv 0$), and the effective Hamiltonian of chiral modes
turns into
\begin{equation}
h_{\mathrm{chiral}}=v\cdot k+i\gamma =v\cdot (k-ik_{0}),  \label{1}
\end{equation}%
where $k_{0}=-\gamma /v$. There is a notable aspect of chiral modes that
\emph{dissipation plays the role of an imaginary wave vector}\cite%
{wangxr2020-1}.

In general, chiral modes are realized as
topologically protected edge states on the boundaries of a 2D Chern
insulator. Therefore, topological edge states are chiral modes under PBC
rather than OBC. \emph{Does there exist NHSE for the chiral modes with PBC?} The answer is \emph{Yes}!

To answer the above questions, we consider a non-Hermitian
Hamiltonian for chiral states with inhomogeneous, perturbative dissipation $\gamma(x)$ ($x\in[0,L]$, $\left \vert \gamma(x) \right \vert \ll 1$).
Additionally, we establish PBC as $\psi(0) =\psi(L)$. The wave functions in $x\in [0,L]$ is expressed as
\begin{equation}
\psi_{k}(x)=\frac{1}{\mathcal{N}} e^{ikx}e^{\frac{1}{v}\int_{0}^{x}
[\gamma(x^{\prime})-\bar{\gamma}] dx^{\prime}},  \label{wave_func}
\end{equation}
where $\mathcal{N}$ is a normalized coefficient, $k$ is the given wave number satisfying $k = (2\pi n)/L$ ($n \in \mathbb{Z}$), and $\bar{\gamma}$ is the average value of the integral of inhomogeneous dissipation in real space, given by
\begin{equation}
    \bar{\gamma} = \frac{1}{L}\int_{0}^{L} \gamma(x) dx.
\end{equation}
The energy levels for chiral modes become
\begin{equation}
    E_{\mathrm{chiral}}=v\cdot k+i\bar{\gamma}, \label{energy_con}
\end{equation}
forming a straight line with fixed imaginary part $\bar{\gamma}$ in complex energy space. Eq.\:(\ref{wave_func}) and Eq.\:(\ref{energy_con}) are derived in Appendix \ref{appendix:a}. Before discussing NHCSE, we introduce \emph{global dissipation} and
its topological defects --- \emph{global dissipation's domain wall} (GDDW).

\textit{Definition 1: Global dissipation: Global dissipation is defined as} $\gamma_{\mathrm{global}}(x)=\gamma(x)-\bar{\gamma}$\textit{ where} $\bar{\gamma}$ \textit{is the average value of inhomogeneous dissipation} $\bar{\gamma}=\frac{1}{L}\int_{0}^{L}\gamma (x)dx.$ \textit{The integral path is a closed loop along the boundary of the chiral edge state.}

\textit{Definition 2: Global dissipation's domain walls: (1) If there exists}
$x_0^{+}$ \textit{that satisfies} $\gamma_{global}(x_0^{+}+0^{-})<0$ \textit{%
and} $\gamma_{global}(x_0^{+}+0^{+})>0$\textit{, we denote} $x_0^{+}$
\textit{as an A-type GDDW. (2) If there exists} $x_0^{-}$ \textit{that
satisfies} $\gamma_{global}(x_0^{-}+0^{-})>0$ \textit{and} $%
\gamma_{global}(x_0^{-}+0^{+})<0$\textit{, we denote $x_0^{-}$ as a B-type
GDDW.}

\begin{figure}[tbp]
\includegraphics[width=8.6cm]{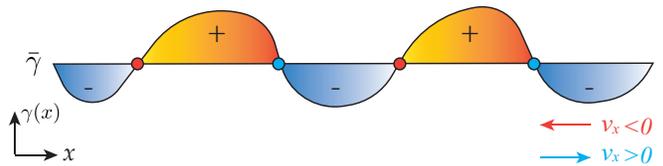}
\caption{The illustration of NHCSE for chiral modes with inhomogeneous
dissipation in the continuous limit. The red regions represent $\protect%
\gamma_{\mathrm{global}}(x)>0$, and the blue regions represent $\protect%
\gamma_{\mathrm{global}}(x)<0$. The red arrow and blue arrow represent the
direction of chiral current $v_x<0,v_x>0$, respectively. The chrial edge
states($v_x<0$) localised at the A-type GDDWs $x_0^{+}$ (red dots), and the
chrial edge states($v_x>0$) localised at the B-type GDDWs $x_0^{-}$ (blue
dots).}
\label{GDDW_con}
\end{figure}

As shown in Fig.\:\ref{GDDW_con}, the A-type GDDWs represent red dots, and
B-type GDDWs represent blue dots. To determine the positions $x_0^{\pm}$ according to definitions 1 and 2, it is necessary to integrate the inhomogeneous dissipation over the closed loop along the boundary of the chiral edge state in order to obtain $\bar{\gamma}$. This is why we refer to it as the \textquotedblleft global \textquotedblright dissipation's domain wall. Furthermore, this highlights that the NHCSE is a type of NHSE specifically under PBC. Subsequently, we present the critical properties of the NHCSE.

\textit{Definition 3: The wave function is defined as localized at} $%
x_0^{\pm}$ \textit{within the interval} $[a,b]$ \textit{if the squared
modulus of the wave function is monotonically increasing in the interval} $[a,x_0^{\pm})$ \textit{and monotonically decreasing in the interval} $%
(x_0^{\pm},b]$ \textit{when} $a<x_0^{\pm}<b$.

\textit{Theorem: If} $\gamma_{global}(x)$ \textit{is a continuous function, then a chiral mode with the negative speed is localized at} $x_0^{+}$ \textit{(within a certain interval) if and only if there exists an A-type GDDW at} $x_0^{+}$\textit{. Furthermore, a chiral mode with the positive speed is localized at} $x_0^{-}$ \textit{(within a certain interval) if and only if there exists a B-type GDDW at} $x_0^{-}$.

We show detailed proofs and examples in Appendix \ref{appendix:b} and \ref{appendix:c}.

This result gives rise to an interesting effect, the NHCSE, signifying the
existence of NHSE for chiral modes under PBC. In particular, NHCSE is quite
different from the usual NHSE (or higher-order NHSE) for bulk states under
OBC. For NHCSE, there \emph{does not} exist point-like topology
configuration in complex spectrum \cite{KZhang2020,okuma2020}. Instead,
in complex spectrum, the energy levels for chiral modes with NHCSE have
a structure of a line rather than a closed loop. It is non-zero global
dissipation $\gamma \mathit{_{\mathrm{global}}}=\gamma (x)-\bar{\gamma}\neq
0 $ in real space rather than a certain \textquotedblleft winding
number\textquotedblright in complex spectrum that plays the role of
an \textquotedblleft order parameter\textquotedblright characterizing
NHCSE. A zero effective dissipation $\gamma_{\mathrm{global}}$ indicates the vanishing of NHCSE. The bigger of $\gamma_{\mathrm{global}}$, the stronger
of NHCSE.

In the following parts, NHCSE is applied to the topologically protected edge
states in 2D Chern insulators with inhomogeneous dissipation. With the help
of this toy lattice model, an additional property of NHCSE --- \emph{anomaly}
will be demonstrated.

\section{Model}
\label{sec.3}
In this section, we consider chiral modes on the boundary of the non-Hermitian
Haldane model with inhomogeneous, perturbative dissipation to describe the
properties of NHCSE. The Haldane model is a typical (Hermitian) model of a
2D Chern insulator on a honeycomb lattice \cite{Haldane1,Haldane2}, of which
the Hamiltonian is
\begin{equation}
\hat{H}_{\mathrm{haldane}}=t_{1}\sum_{\langle ij\rangle }c_{i}^{\dag
}c_{j}+t_{2}\sum_{\langle \langle ij\rangle \rangle }e^{i\phi
_{ij}}c_{i}^{\dag }c_{j},
\end{equation}%
where $c_{i}^{\dagger }$ and $c_{i}$ are creation and annihilation operators
for a particle at the $i$-th site. $\langle i,j\rangle $ and $%
\langle \langle i,j\rangle \rangle $ denote the nearest-neighbor (NN)
hopping and the next-nearest-neighbor (NNN) hopping, and $t_{1}$ and $t_{2}$
are the strength of NN hopping and NNN hopping, respectively. $e^{i\phi
_{ij}}$ is a complex phase of the NNN hopping, and we set the direction of
the positive phase as clockwise $(|\phi _{ij}|=\frac{\pi }{2})$. In this
paper, we set $t_{1}$ to be a unit, and $t_{2}$ to be constant, i.e., $%
t_{2}\equiv 0.2t_{1}$.

The topological characterization of the Haldane model is captured by the
Chern number $Q$. Due to the existence of nonzero Chern number $Q=1$, there
exist topological edge states propagating along the system's edge \cite%
{Haldane1}. For a system with PBC along $x$ direction, and an OBC along $y$ direction with zigzag edges, the effective (single-body) Hamiltonian of topological edge
states $h_{\mathrm{edge}}(k_{x})$ becomes
\begin{equation}
h_{\mathrm{edge}}(k_{x})=\pm v_{\mathrm{eff}}\cdot \sin k_{x},
\end{equation}
where $v_{\mathrm{eff}}$ characterizes the speed of topological edge states and can be described as
\cite{huang_2012}
\begin{equation}
    v_{\mathrm{eff}}=\frac{6t_{1}t_{2}}{\sqrt{%
t_{1}^{2}+8t_{2}^{2}(1-\cos k_{x})}}.
\end{equation}

According to $h_{\mathrm{edge}}(k_{x}),$ we say that the
topological edge states of such a Chern insulator on a honeycomb lattice are
chiral modes.

To illustrate the NHCSE, we consider a non-Hermitian Haldane model with
inhomogeneous, perturbative dissipation. The Hamiltonian
becomes
\begin{equation}
\hat{H}_{\mathrm{total}}=\hat{H}_{\mathrm{haldane}}+\delta \hat{H}_{\mathrm{D%
}},  \label{model_ham}
\end{equation}%
where $\delta \hat{H}_{\mathrm{D}}$ represents the term associated with dissipation. In the subsequent sections, we investigate the NHCSE for three cases of $\delta \hat{H}_{\mathrm{D}}$: (a) Bulk dissipation (Section \ref{case(a)}), (b) Dissipation on a single outermost zigzag edge (Section \ref{case(b)}), and (c) Staggered on-site gain/loss (Section \ref{sec.6}).

\section{Non-Hermitian chiral skin effect for bulk dissipation DWs}

\label{case(a)} Now, we consider bulk's dissipation on the Haldane model,
i.e.,
\begin{equation}
    \delta \hat{H}_{\mathrm{D}}=i\sum \limits_{\mathrm{bulk}}\gamma _{i}c_{i}^{\dag
}c_{i}.
\end{equation}
where $\gamma_{i}$ represents the strength of dissipation at the $i$-th lattice site among all lattice sites. Firstly, we discuss the Haldane model with uniform bulk dissipation, $\gamma
_{i}=\gamma $. When $\gamma \neq 0$, the imaginary part of energy levels $%
E(k_{x},y)$ in complex spectrum have a global, uniform, shift,
\begin{equation}
\text{Im}E(k_{x},y)=0\rightarrow \text{Im}E(k_{x},y)\equiv \gamma .
\end{equation}%
The dispersion relations of both edge modes are given by
\begin{equation}
h_{\mathrm{edge}}(k_{x})=\pm v_{\mathrm{eff}} \cdot \sin k_{x} + i\gamma .
\label{bulk_effective ham}
\end{equation}
In continuous limit, near $k_{x}=\pi$, $h_{\mathrm{edge}}(k_{x})$ is
reduced into Eq.\:(\ref{1}), i.e.,
\begin{equation}
h_{\mathrm{edge}}(k_{x})\simeq \pm v_{\mathrm{eff}}\cdot (\Delta k\pm
i\gamma /v_{\mathrm{eff}})
\end{equation}%
where $\Delta k=k_{x}-\pi$.

\begin{figure}[tbp]
\includegraphics[width=9.8cm]{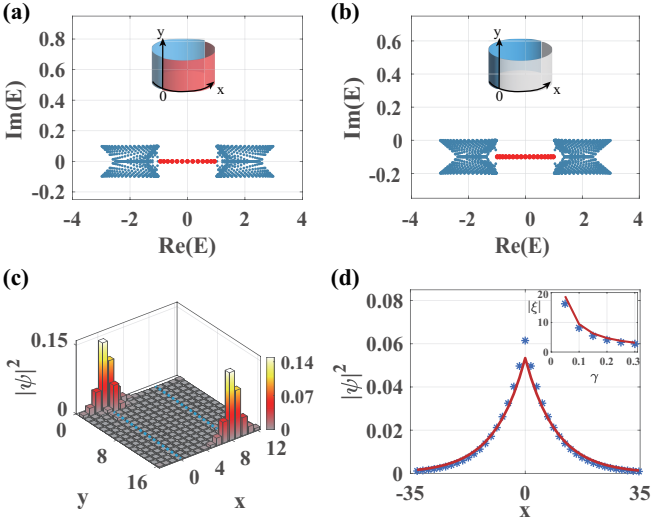}
\centering
\caption{Non-Hermitian Haldane model in the cylinder geometry ($x$PBC/$y$%
OBC) with GDDWs in bulk. (a)-(b) The complex spectra with GDDWs (a)$\gamma_L=-0.1t_{1}, \protect\gamma_R=0.1t_{1}$ and (b)$\protect\gamma%
_L=-0.2t_{1}, \protect\gamma_R=0$. The blue and red dots represent the
energy levels of bulk states and topological edge states, respectively. (c) Particle distribution of the topological edge states. The dissipation is set as $\protect\gamma_{R}(0<x<8)=0.2t_{1}$, $%
\protect\gamma_{L}(-4<x<0,8<x<12)=-0.2t_{1}$. (d)Comparison between
numerical results (blue stars) and theoretical predictions (red line) of the
particle distribution of a topological edge state at the edge ($y=0$). The
inset displays the localization length $\protect\xi$ for varying strengths
of $\protect\gamma$. The numerical results are obtained from the
Non-Hermitian Haldane model, while the theoretical results are derived from
Eq.\:(\ref{local_lenghth_bulk}). Both wave functions are normalized
along the $x $ direction.}
\label{fig1}
\end{figure}

Next, to verify the existence of NHCSE, we discuss the Haldane model in the cylinder geometry ($x$PBC/$y$OBC) with inhomogeneous bulk dissipation by considering a pair of GDDWs in bulk. As shown in Fig.\:\ref{fig1}(a) and (b), the cases where there is a difference in the bulk's dissipation between the left and right regions are investigated. In both model, we show the existence of NHCSE --- the energy levels of
topological edge states become a line in the complex energy spectrum [Fig.\:\ref%
{fig1}(a) and (b)] and the topological edge states accumulate around the
GDDWs [Fig.\:\ref{fig1}(c)]. In addition, the numerical results match our
analytical prediction based on Eq.\:(\ref{bulk_effective ham}) [Fig.\:\ref%
{fig1}(d)]. The localization length $\xi $ becomes
\begin{equation}
\xi= \frac{v_{\mathrm{eff}}}{\gamma}, \label{local_lenghth_bulk}
\end{equation}
and the specific derivations can be found in Appendix \ref{appendix:c}.

In summary, such NHCSE can never be characterized by the certain point-like topology configuration in complex spectrum or corresponding
\textquotedblleft winding number\textquotedblright! Instead, it is global
dissipation $\gamma_{\mathrm{global}}$ that can characterize the strength of
NHCSE.\

\section{Non-Hermitian chiral skin effect for edge dissipation DWs}

\label{case(b)} We then consider the effect of inhomogeneous dissipation on the single outermost zigzag edge, i.e.,
\begin{equation}
    \delta \hat{H}_{\mathrm{D}}=i\sum_{\mathrm{%
lower-edge}}\gamma _{i}c_{i}^{\dag }c_{i}.
\end{equation}
\begin{figure}[tbp]
\includegraphics[width=9.8cm]{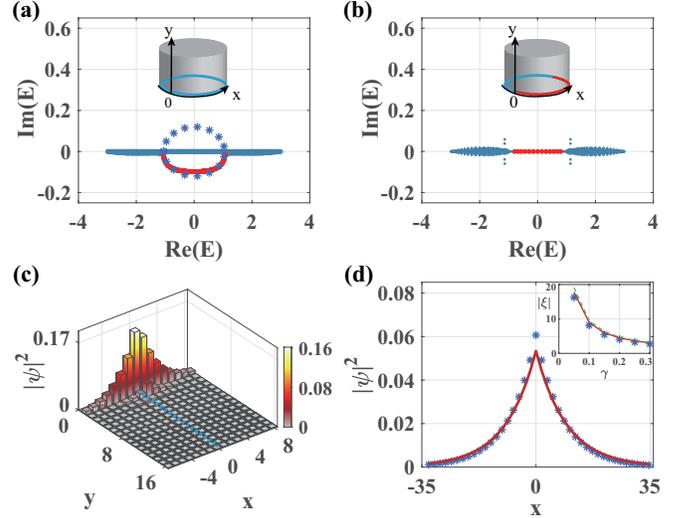}
\caption{Non-Hermitian Haldane model in the cylinder geometry ($x$PBC/$y$OBC)
with dissipation on a single outermost zigzag edge ($y=0$). (a)-(b)
Complex spectra with (a)uniform dissipation $\protect\gamma=-0.1t_{1}$ and
(b) GDDW ($\protect\gamma_L=-0.1t_{1}, \protect\gamma_R=0.1t_{1}$). Blue
dots, red dots, and cobalt blue stars represent energy levels of bulk
states, topological edge states in the non-Hermitian Haldane model, and
energy levels of effective 1D Hatano-Nelson model [Eq.\:(\protect\ref%
{edge_effective_ham})], respectively. (c) Particle distribution of topological edge states. The GDDW is set as $\gamma _{R}(x>0)=0.2t_{1}$,
$\gamma _{L}(x<0)=-0.2t_{1}$. (d) Comparison between numerical and
theoretical results for the particle distribution of topological edge states
at $y=0$. The inset shows the localization length $\protect\xi$ for varying
dissipation strengths $\protect\gamma$. Blue stars, red lines, and green
dash lines represent numerical results for the non-Hermitian Haldane model,
theoretical results for the half Hatano-Nelson model[Eq.\:(\ref{local_hn})], and continuous limits[Eq.\:(\protect\ref{local_lenghth_bulk})]. Both wave functions are
normalized along the $x$ direction.}
\label{fig2}
\end{figure}
\subsection{Effective Model for Topological Edge
States: Half Hatano-Nelson model}
First, we point out that \emph{the effective model for topological edge
states in this condition is Half Hatano-Nelson model that could exactly
characterize NHCSE}. The effective Hamiltonian for the topological edge states is given by
\begin{equation}
h_{\mathrm{edge}}(k_{x})=\pm v_{\mathrm{eff}}\cdot \sin k_{x}+i\gamma_{\mathrm{eff}},
\end{equation}
where $k_{x}=\pi +\delta k$, and $\gamma_{\mathrm{eff}}$ represents the effective dissipation, defined as
\begin{equation}
\gamma_{\mathrm{eff}}=\text{Im}\left\langle \psi_{\mathrm{edge}}\right\vert \hat{H}_{\mathrm{total}}\left\vert \psi_{\mathrm{edge}}\right\rangle.
\end{equation}
Here, $\psi _{\mathrm{edge}}$ denotes the wave functions of the chiral modes.

To accurately characterize NHCSE, we obtain the effective Hamiltonian for
topological edge states by data fitting the effective dissipation for a
uniform case $\gamma _{i}=\gamma $. Now, we have
\begin{equation}
\gamma _{\mathrm{eff}}(k_{x})=\sum_{n}a_{n}\cos nk_{x}.
\end{equation}%
For example, in the case of $\gamma =0.1t_{1},$ we have $a_{1}\simeq
0.12t_{1}$, $a_{2}\simeq 0.023t_{1}$. As a result, the effective Hamiltonian
for chiral edge states $\hat{h}_{\mathrm{edge}}(\hat{k}_{x})$ with uniform
edge dissipation becomes half of a 1D Hatano-Nelson model described by
\begin{align}
h_{\mathrm{edge}}(k_{x})& =h_{\mathrm{half-HN}}(k_{x})\simeq v_{\mathrm{eff}%
}\sin k_{x}+i\gamma _{\mathrm{eff}}(k_{x}),  \notag \\
k_{x}& \in (-\pi /2,\pi /2],
\label{edge_effective_ham}
\end{align}%
where $\gamma _{\mathrm{eff}}(k_{x})=a_{1}\cos k_{x}+a_{2}\cos 2k_{x}$.
For clarity, we map Eq.(\ref{edge_effective_ham}) to the usual 1D generalized Hatano-Nelson model, we have

\begin{align}
\hat{H}_{\mathrm{HN}} &= \sum_{i}(t_{L}c_{i}^{\dag}c_{i+1}+t_{R}c_{i+1}^{\dag
}c_{i}) \\ \nonumber
&= \sum_{k_{x}\in(-\pi,\pi)}a_{k_{x}}^{\dag}(v_{\mathrm{eff}}\sin
k_{x}+i\zeta\cos k_{x})a_{k_{x}},
\end{align}
where $t_{L}=(v_{\mathrm{eff}}-\zeta)/2$ and $t_{R}=(v_{\mathrm{eff}}%
+\zeta)/2$. According to the 1D Hatano-Nelson model, the localization length in
non-Hermitian Haldane model is described by
\begin{equation}
\xi= \frac{c}{2 \ln{\frac{t_{R}}{t_{L}}}}, \label{local_hn}
\end{equation}
where $c=\sqrt{3}a$ and $a$ is lattice constant of Haldane model.

The characteristics of the complex spectrum under edge dissipation are discussed. The numerical results given by the non-Hermitian Haldane model match our effective model (half of the Hatano-Nelson). As shown in Fig.\:\ref{fig2}(a), the red line represents the numerical energy levels of the topological edge states under PBC and the cobalt blue stars represent the numerical energy levels of the 1D Hatano-Nelson model. By comparing the energy levels, it is found that the effective model of chiral modes at $y=0$ is only half of a 1D Hatano-Nelson model, which is the negative imaginary parts of the energy levels of the 1D Hatano-Nelson model. Moreover, the numerical results of the particle distribution of topological edge states matches our analytical prediction based on Eq.\:(\ref{edge_effective_ham}) [Fig.\:\ref{fig2}(d)]. The inset in Fig.\:\ref{fig2}(d) illustrates that the numerical results for the localization length $\xi$ correspond to the localization length of the half Hatano-Nelson model [Eq.\:(\ref{local_hn})].

The localization behavior of topological edge states under edge dissipation is investigated using the NHCSE. Firstly, we introduce uniform dissipation on the outermost edge at $y=0$ with $\gamma_i=-0.1t_1$, as shown in Fig.\:\ref{fig2}(a). The complex spectrum reveals that the chiral edge states on the dissipative edge exhibit non-zero effective dissipation. Next, we consider an asymmetric dissipation configuration on the outermost edge ($y=0$), where one half of the system has dissipation $\gamma(x<0)=-0.1t_1$, and the other half has dissipation $\gamma(x>0)=0.1t_1$, as depicted in Fig.\:\ref{fig2}(b). By utilizing the NHCSE theory, we predict the emergence of A-type GDDWs on the dissipative edge at $x=0$, which are associated with the direction of chiral current $v_x<0$ at the edge $y=0$. The numerical results of the particle distribution, as depicted in Fig.\ref{fig2}(c), confirm the localization of chiral edge states at the A-type GDDWs, thus validating the theoretical predictions.

Therefore, NHCSE can be considered as \textquotedblleft \emph{anomaly}
\textquotedblright\ NHSE and only is realized on the boundaries (or certain DWs) of a 2D Chern insulator.

\subsection{The Non-Local Effect of NHCSE}
\begin{figure}[tbp]
\includegraphics[width=8.6cm]{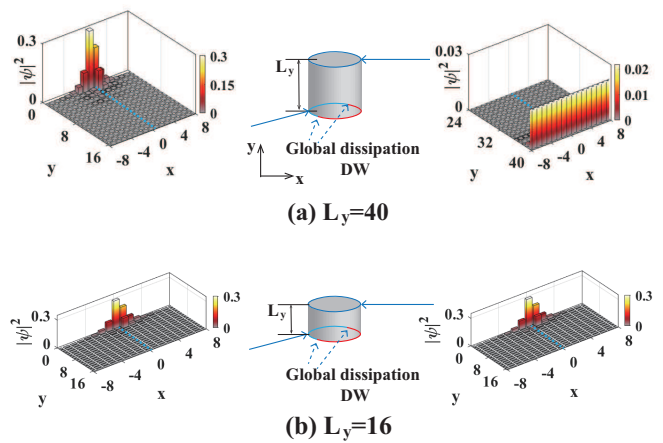}
\caption{The non-local NHCSE. The nanoribbons of the Haldane model with a pair of GDDW on a single outermost zigzag edge ($y=0$) ($\protect\gamma _{R}(x>0)=0.5t_{1},$ $\protect\gamma _{L}(x<0)=-0.5t_{1}$). (a) The topological edge states on edge ($y=0$) are localized on the GDDW, and
the other on edge ($y=L_y$) are extended on its corresponding
edge. (b) The topological edge states on both edges ($y=0$, $y=L_y$) are localized on the GDDW. }
\label{fig3}
\end{figure}

Second, we discuss the interplay between the topological edge states on both
edges (the edge with dissipation and that without) and show the non-local
effect of NHCSE. We consider a non-Hermitian Haldane model on a strip with
zigzag edges, of which one edge has dissipation ($y=0$) and the other has
not ($y=L_{y}$). The distance between two edges is set to be $L_{y}$, as
shown in Fig.\:\ref{fig3}(a).

We consider the GDDWs on the edge ($y=0$). In the limit of $L_{y}\rightarrow
\infty $ (for example, $L_{y}=40$), the results look trivial --- due to
NHCSE, the topological edge states on the edge with GDDWs become localized
on GDDWs; the topological edge states on the edge without dissipation are
extended, as shown in Fig.\:\ref{fig3}(a). In the scenario where $L_{y}$ has
a small value (for example, $L_{y}=16$), an unusual occurrence takes place
whereby the topological edge states on the edge with dissipation as well as
the ones on the edge without it \emph{both} become localized on the same
GDDWs. To emphasize the results, we call it \emph{non-local NHCSE}, as shown in Fig.\:\ref{fig3}(b). In the future, we will study the mechanism for non-local NHCSE and try to give a reasonable answer.

\begin{figure*}[htbp]
\includegraphics[width=17.2cm]{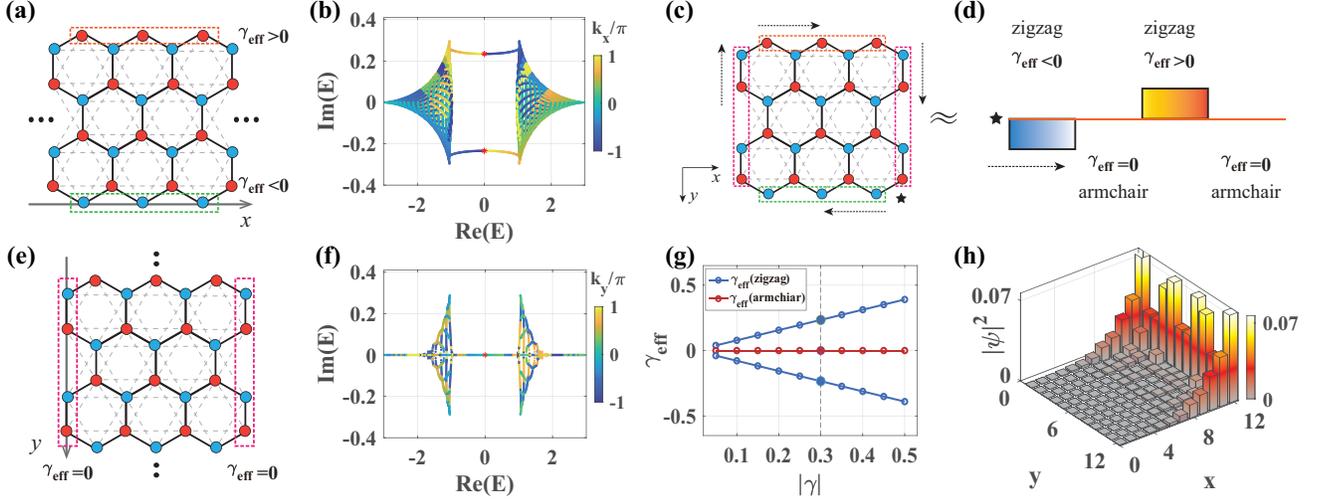}
\caption{Illustration of NHCSE with staggered gain/loss. (a),(e) Schematic
diagram of the non-Hermitian Haldane model with zigzag edges(a) and armchair edges(e). The orange boxes represent regions of effective gain, the green
boxes represent regions of effective loss, and the magenta boxes represent
regions without dissipation. (b),(f) Energy levels in the complex spectrum for zigzag edges(b) and armchair edges(f). The red stars
represent the effective dissipation on each edge. (c)Schematic diagram of the
non-Hermitian Haldane model under OBC with zigzag and armchair edges. The
dotted arrows indicate the direction of the chiral current. (d) Schematic
diagram multi-GDDWs with the direction of the chiral current $v>0$, which is equivalent to (c). (g) Variation of effective edge dissipation on the zigzag and
armchair edges for varying strengths of $\protect\gamma$. (h)Spatial
distribution of particles in the topological edge states corresponding to
the configuration in (c). The strength of gain/loss is $|\protect\gamma|=0.3t_{1}$.
}
\label{fig.app_zigzag_armchair}
\end{figure*}

\section{Application: Mechanism of Hybrid Skin-Topological Effect in Chern
Insulators}
\label{sec.6}
In this section, we explain the mechanism of the gain/loss-induced hybrid skin-topological effect and demonstrate that this effect can be understood in terms of multiple GDDWs for topological edge states in the theory of NHCSE. Specifically, we focus on the non-Hermitian Haldane model and the staggered on-site gain/loss can be expressed as
\begin{equation}
\delta\hat{H}_{\mathrm{D}}=i\gamma\sum_{i}c_{A,i}^{\dag}c_{A,i}-i%
\gamma\sum_{i}c_{B,i}^{\dag}c_{B,i},
\end{equation}
where A/B are the two sublattice sites in each subcell and $\gamma$ denotes the strength of dissipation. It holds significant importance as it is firstly proposed the gain/loss-induced hybrid skin-topological effect\cite%
{HLiu2022}.

In our study, NHCSE is the origin of the hybrid skin-topological effect,
which arises from the presence of global effective dissipation at the
boundaries. Next, we will demonstrate that staggered on-site gain/loss
throughout the entire system can be equivalently regarded as effective
dissipation at the boundaries for topological edge states.

To obtain the effective dissipation at each boundary, we consider two
cylindrical Haldane systems in PBC along the $x$ direction and OBC along the $y$ direction($x$PBC/$y$obc) with zigzag edges:[Fig.\:\ref{fig.app_zigzag_armchair}(a)], and $x$OBC/$y$PBC with armchair edges[Fig.\:\ref{fig.app_zigzag_armchair}(e)]. The effective edge dissipation with zigzag edges for topological edge states is given by
\begin{align}
    \gamma _{\mathrm{eff}}^{\mathrm{zigzag}}=&\text{Im}\left\langle \psi _{%
\mathrm{edge}}(k_{x})\right\vert (i\gamma\sum_{i}c_{A,i}^{\dag}c_{A,i}\\ \nonumber
&-i\gamma\sum_{i}c_{B,i}^{\dag}c_{B,i})\left\vert \psi _{_{%
\mathrm{edge}}}(k_{x})\right\rangle,
\end{align}
and the effective edge dissipation with armchair edges for topological edge states is given by
\begin{align}
    \gamma _{\mathrm{eff}}^{\mathrm{armchair}}=&\text{Im}\left\langle \psi _{\mathrm{edge}}(k_{y})\right\vert (i\gamma\sum_{i}c_{A,i}^{\dag}c_{A,i}\\ \nonumber
&-i\gamma\sum_{i}c_{B,i}^{\dag}c_{B,i})\left\vert \psi _{_{%
\mathrm{edge}}}(k_{y})\right\rangle,
\end{align}
where $\psi _{\mathrm{edge}}$ denote topological edge states. As shown in
Fig.\ref{fig.app_zigzag_armchair}(b) and (f), We plot the complex energy
spectrum for the case of zigzag edges[Fig.\ref{fig.app_zigzag_armchair}(b)]
and armchair edges[Fig.\ref{fig.app_zigzag_armchair}(f)]. In the energy
spectrum, the effective dissipation associated with chiral modes on the two
edges are indicated by red stars. For the topological edge states on the
zigzag edge, where there is no symmetry between the A and B sub-lattices,
the contributions do not cancel out, leading to a non-zero effective
dissipation. However, on the armchair edge, the contributions from the A and
B sub-lattices cancel out due to their symmetry, resulting in zero effective
dissipation. The effective dissipation on the zigzag and armchair edges[Fig.%
\ref{fig.app_zigzag_armchair}(a) and (e)] as a function of $\gamma $ are
depicted in Fig.\:\ref{fig.app_zigzag_armchair}(g). This finding illustrates
that the skin effect induced by the global staggered on-site gain/loss can
be effectively described as a theoretical model of NHCSE, which is
determined by effective dissipation at the boundaries.

As previously discussed, the hybrid skin-topological effect induced by
staggered gain/loss is determined by the effective dissipation at the
boundaries. Due to chiral edge currents propagating along the geometric
boundaries of the system under OBC, it suggests that chiral edge states are
determined by the effective dissipation on the system's geometric
boundaries. To explain the mechanism of the hybrid skin-topological effect, we
map the hybrid skin-topological effect under OBC [Fig.\ref%
{fig.app_zigzag_armchair}(c)] to the multi-GDDWs issue in the framework of
NHCSE[Fig.\ref{fig.app_zigzag_armchair}(d)]. Here, we provide a specific
system with rectangular borders to illustrate the skin effect of
topological edge states.

In the case of OBC with rectangular borders, the effective dissipation along the four edges can be obtained as shown in Fig.\:\ref{fig.app_zigzag_armchair}(g), corresponding to $-\gamma_{\mathrm{eff}}^{\mathrm{zigzag}}, 0, \gamma_{\mathrm{eff}}^{\mathrm{zigzag}}, 0$, respectively. This mapping successfully transforms the case of OBC with rectangular borders in Fig.\:\ref{fig.app_zigzag_armchair}(c) into the multi-GDDWs model depicted in Fig.\:\ref{fig.app_zigzag_armchair}(d). By analyzing the multi-GDDWs model of NHCSE, we explain the hybrid skin-topological effect under OBC with rectangular borders. In Fig.\:\ref{fig.app_zigzag_armchair}(d), the direction of chiral current is positive ($v>0$), based on Eq. (\ref{wave_func}) and continuity condition at each GDDW, the wave function increases along the $x$-axis in the region where $\gamma_{\mathrm{eff}}-\bar\gamma>0$, decreases in the region where $\gamma_{\mathrm{eff}}-\bar\gamma<0$, and extended in the region where $\gamma_{\mathrm{eff}}-\bar\gamma=0$. In addition, Fig. \ref{fig.app_zigzag_armchair}(h) provides evidence supporting our theory. It shows the extension of chiral edge states at the armchair boundary, where the effective dissipation is zero. Conversely, at the zigzag boundary, the chiral edge states exhibit localization in a specific direction, indicative of the skin effect. Further detailed derivations for the solutions of the multi-GDDWs model can be found in Appendix \ref{appendix:d}.

In summary, we conclude that NHCSE is the mechanism behind the hybrid skin-topological effect in Chern insulators, distinguishing it from 1D chain NHSE or higher-order NHSE. Our findings demonstrate that hybrid skin-topological effect can be observed by introducing inhomogeneous effective dissipation at the system boundaries.

\section{Realization of an electric circuit in Haldane model with Global dissipation domain wall}
\label{sec.7}
In this section, we present a circuit design that implements the non-Hermitian Haldane model with global dissipation domain walls.

\begin{figure}[ptb]
\includegraphics[width=8.6cm]{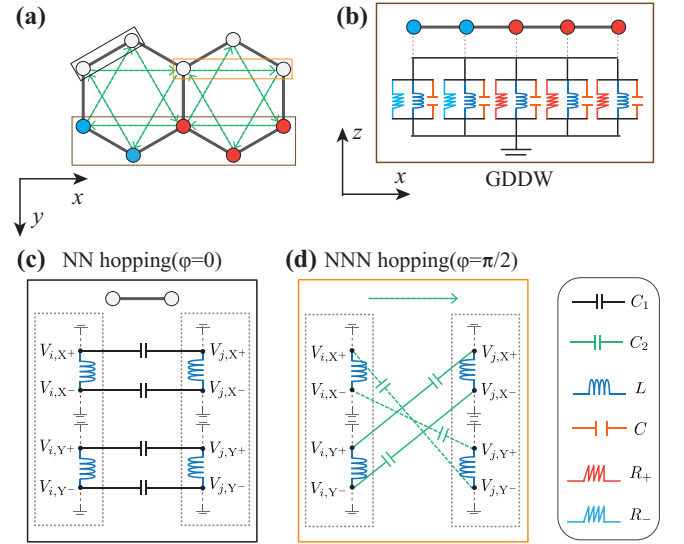}
\caption{The schematic of the designed electric circuit. (a)Illustration of non-Hermitian Haldane model. The brown box indicates the GDDW shown in (b), the black box indicates the NN hoping shown in (c), and the orange box indicates the NNN hoping shown in (d). (b) The circuit implementation of the GDDW. (c) Schematic of the circuit implementation of the NN hopping. (d) Schematic of the circuit implementation of the NNN hopping,}
\label{ele_circuit}
\end{figure}

Our experimental platform utilizes an LC circuit combined with resistors ($R_{-}$) and negative resistors (INIC) ($R_{+}$) to observe NHCSE in the non-Hermitian Haldane model with GDDW. To map the Haldane model with GDDW onto the electric circuit, we introduce various types of couplings\cite{Zhang_2022,cong} and on-site gain/loss\cite{negative_resistance}. Three charts in Fig.\:\ref{ele_circuit} illustrate the circuit implementations of nearest-neighbor (NN) hopping[Fig.\ref{ele_circuit}(c)], next-nearest-neighbor (NNN) hopping[Fig.\ref{ele_circuit}(d)], and GDDWs[Fig.\ref{ele_circuit}(b)]. In Fig.\ref{ele_circuit}(c) and (d), the gray dotted box represents a lattice site in a tight-binding model, containing two inductors (X, Y). The voltages between the inductors, denoted as $U_{X}$ and $U_{Y}$, are used to define the variables $U_{\uparrow,\downarrow}=U_{X}\pm iU_{Y}$. By applying Kirchhoff's law and utilizing this notation, we derive the eigenequation for the spin-up and spin-down states
\begin{align}
E\left[
\begin{array}{l}
U_{\mathbf{k},\uparrow}^{A} \\
U_{\mathbf{k},\uparrow}^{B}%
\end{array}
\right] =\left[
\begin{array}{ll}
p_{k}(\varphi)-M_{A} & T_{\mathbf{k}} \\
T_{\mathbf{k}}^{\ast} & p_{k}(-\varphi)-M_{B}%
\end{array}
\right] \left[
\begin{array}{l}
U_{\mathbf{k},\uparrow}^{A} \\
U_{\mathbf{k},\uparrow}^{B}%
\end{array}
\right],   \label{18}
\end{align}
where the energy is characterised by $E=3t_{1}+6t_{2}-\frac{2w_{0}^{2}}{w^{2}}$, the NNN hopping is characterised by $p_{k}(\varphi )=2t_{2}\left[ \cos (k\cdot v_{1}+\varphi )+\cos (k\cdot v_{2}+\varphi )+\cos (k\cdot v_{3}+\varphi )\right]$, the NN hopping is characterised by $T_{\mathbf{k}}=t_{1}(e^{i\mathbf{ke}_{1}}+e^{i\mathbf{ke}_{2}}+e^{i\mathbf{ke}_{3}})$, and the on-site term is characterised by $%
M_{A(B)}=(C_{g}-\frac{1}{w^{2}L_{g}}-\frac{i}{wR_{g,A(B)}})/C$. The specific derivations can be found in Appendix \ref{appendix:e}.
For convenience, the grounding capacitance and inductance are set as $C_{g}=C$, $L_{g}=L$, and capacitance representing coupling is set as $C_{1}=t_{1}C$, $C_{2}=t_{2}C$.

The eigenfrequency of circuit is $\omega _{0}=1/(LC)^{1/2}$, and Eq.\:(\ref{18}) is the eigefunction of the non-Hermitian Haldane model. Notably, the on-site dissipation is described as
\begin{equation}
    i\gamma _{A(B)}=-M_{A(B)}=-\frac{i}{R_{g,A(B)}}\sqrt{\frac{L}{C}}.
\end{equation}
It's to be noted that $R_{g, A(B)}$ can be negative by
using INIC, which corresponded to gain in the non-Hermitian Haldane model. We show more details in Appendix \ref{appendix:e}.

Let us now consider the electric circuit that has PBC along the $x$ direction, and an OBC along the $y$ direction with zigzag edges, the NN and NNN coupling is grounded at $y=0$ and $y=L$. To achieve a pair of GDDWs, we utilize two approaches. In the first approach, we place half of the resistors with a value of $R_1$ and half of the resistors with a value of $R_2$ at the corresponding grid points on the boundary. Alternatively, in the second approach, we position half of the resistors with a value of $R_1$ and an equal number of the same type of INIC at the corresponding grid points on the boundary. These strategies enable us to create the GDDW and facilitate the desired dissipation distribution. In semi-infinite case, the voltage distribution at the eigenfrequency $\omega_{0}$ is measured to
obtain the corresponding eigenvalue $E$, and the eigenstate $[U_{\mathbf{k},\uparrow }^{A},U_{\mathbf{k},\uparrow}^{B}]^{T}$. These simulation experiments are convenient and clear to observe the NHCSE phenomenon.

\section{Conclusion}
\label{sec.8}
\begin{figure}[htbp]
\includegraphics[width=8.6cm]{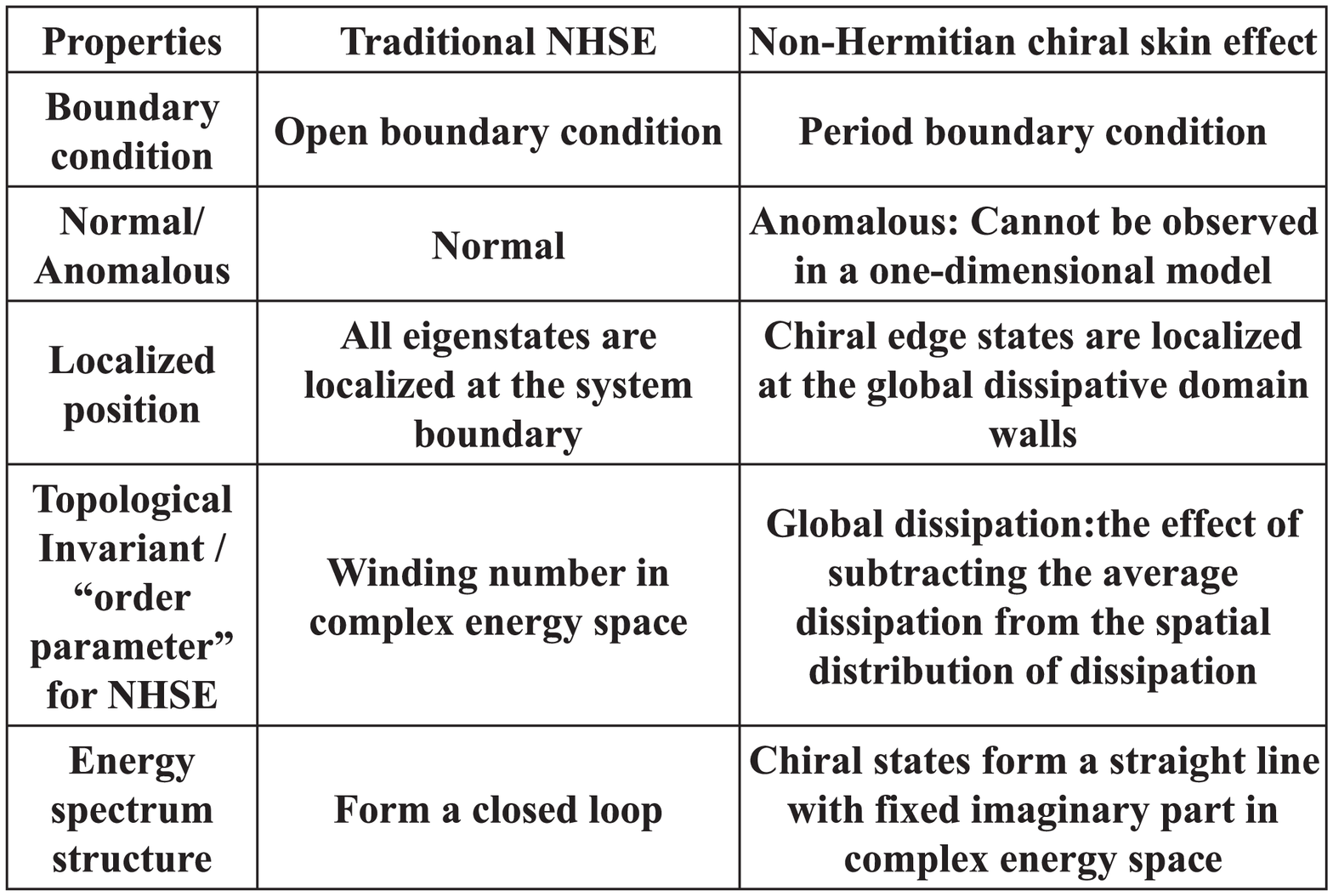}
\caption{The schematic of the difference between traditional NHSE and NHCSE.}
\label{difference_nhse_nhcse}
\end{figure}

In the end, we answer the three questions at the beginning: (\rmnum{1}) We
found a fundamental difference between the NHSE for bulk states and NHCSE
for topological edge states with inhomogeneous dissipation. One key point is that NHCSE is a \textquotedblleft NHSE \textquotedblright under PBC rather than OBC. The chiral modes localize around GDDWs -- a special topological defect of global dissipation rather than the boundaries of the systems. Another key point is \textquotedblleft anomaly\textquotedblright. For example, the NHCSE for topological edge states is characterized by half of the Hatano-Nelson model rather than a whole 1D tight-binding lattice model. The specific distinctions are shown in Fig.\:\ref{difference_nhse_nhcse}. (\rmnum{2}) We found that without point-like topological configuration, NHCSE is characterized by the \textquotedblleft global dissipation\textquotedblright \ $\gamma _{\mathrm{%
global}}$ rather than a certain \textquotedblleft winding
number\textquotedblright . (\rmnum{3}) We have a deeper understanding of the hybrid skin-topological effect in Chern insulators with the help of NHCSE in multiple GDDWs for topological edge states. In summary, we conclude that
\begin{align}
& \text{Chiral modes + Inhomogeneous dissipation}  \notag \\
& \rightarrow \text{Non-Hermitian chiral skin effect.}  \notag
\end{align}

In the future, we will try to systematically understand the interplay of
non-Hermitian physics and topological quantum states by studying the NHCSE
for topological edge states in other types of topological quantum states,
such as higher-order Chern insulators, topological superconductors, and
topological semi-metals.
\section*{ACKNOWLEDGMENTS}
This work was supported by the Natural Science Foundation of China (Grants
No. 11974053 and No. 12174030 ). We are grateful to Zhong Wang, Chen Fang,
Lin-Hu Li, Wei Yi, Gao-yong Sun, Ya-jie Wu, Can Wang, Fei Yang, Xian-qi Tong, and Yue Hu for helpful discussions that contributed to clarifying some aspects related to the
present work.
\appendix
\section{The Wave Functions of Chiral Edge States with Dissipation $\protect%
\gamma(x)$}
\label{appendix:a}
The effective edge model of the topological edge states is given by the
Hamiltonian%
\begin{equation}
\hat{H}= v k + i\gamma(x),
\end{equation}
or%
\begin{equation}
\hat{H}=-iv \frac{d}{dx} + i\gamma(x),
\end{equation}
where $x \in[0,L]$, $v$ is the velocity of the edge states, $k$ is the
wavenumber, and $\gamma(x)$ represents the dissipation as a function of
position $x$.

To obtain the wave function of the topological edge states, we consider the
stationary Schrodinger equation%
\begin{equation}
[-iv \frac{d}{dx} + i\gamma(x) ]\psi(x)=E \psi(x),
\end{equation}
where $E$ is the energy of the system. Solving this equation, we obtain the wave function%
\begin{equation}
\psi(x)=\frac{1}{\mathcal{N}}e^{i\frac{E}{v}x}e^{\frac{1}{v}\int\gamma(x)
dx},
\end{equation}
where $\mathcal{N}$ is the normalization factor. Here, $E$ can be any
complex number.

To determine the energy $E$, we impose PBC $%
\psi(0)=\psi(L)$, which yield%
\begin{equation}
i\frac{E}{v}L+\frac{1}{v}\int_{0}^{L} \gamma(x) dx =i2\pi n,
\end{equation} where $n \in\mathbb{Z}$. This equation gives us the energy of the edge states, which can be expressed as%
\begin{equation}
E= vk + i\bar{\gamma},
\end{equation}
where $k= \frac{2\pi n}{L}$, and $\bar{\gamma}=\frac{1}{L}\int_{0}^{L}
\gamma(x) dx$. Substituting this expression for $E$ into the wave function,
we obtain

\begin{equation}
\psi_{k}(x)=\frac{1}{\mathcal{N}} e^{ikx}e^{\frac{1}{v}\int_{0}^{x}
[\gamma(x)-\bar{\gamma}] dx}.  \label{wave_func_sp}
\end{equation}
This equation describes the wave function of the edge states for a given
wavenumber $k$. We define global dissipation as $\gamma _{\mathrm{global}%
}(x)=\gamma (x)-\bar{\gamma}$, where $\bar{\gamma}$ is the average value of
inhomogeneous dissipation $\bar{\gamma}=\frac{1}{L}\int_{0}^{L}\gamma (x)dx$.

\section{Proof of the Relationship between Wave Function Localization and
the Existence of GDDWs}
\label{appendix:b}
\textit{Definition 1:} $x_0$ \textit{is called an A-type GDDW on the interval} $[a,b]$\textit{. We always make} $a$ \textit{and} $b$ \textit{belong to} $[0, L]$ \textit{by changing the initial point, if there exist three positive real numbers, }$a<x_0<b$\textit{, such that for any} $x_L$ \textit{satisfying} $a<x_L<x_0$\textit{, }$\gamma_{global}(x_L)<0$\textit{, and for any }$x_R$ \textit{satisfying} $x_0<x_R<b$\textit{, }$\gamma_{global}(x_R)>0$\textit{.}

\textit{Definition 2:} $x_0$ \textit{is called a B-type GDDW on the interval} $[a,b]$\textit{. We always make} $a$ \textit{and} $b$ \textit{belong to} $[0, L]$ \textit{by changing the initial point, if there exist three positive real numbers,} $a<x_0<b$\textit{, such that for any} $x_L$ \textit{satisfying} $a<x_L<x_0$\textit{, }$\gamma_{global}(x_L)>0$\textit{, and for any} $x_R$ \textit{satisfying} $x_0<x_R<b$\textit{,} $\gamma_{global}(x_R)<0$\textit{.}

\textit{Definition 3:} \textit{The wave function is said to be localized at }$x_0$ \textit{within the interval }$[a,b]$ \textit{if the squared modulus of the wave function is monotonically increasing in the interval} $[a, x_0)$ \textit{and monotonically decreasing in the interval }$(x_0,b]$ \textit{when} $a<x_0<b$\textit{.}

\textit{Theorem:} \textit{If} $\gamma_{global}(x)$ \textit{is a continuous function, then a chiral mode with a negative speed is localized at} $x_0$ \textit{(within a certain interval) if and only if there exists a type-A GDDW at} $x_0$\textit{. Furthermore, a chiral mode with a positive speed is localized at} $x_0$ \textit{(within a certain interval) if and only if there exists a type-B GDDW at} $x_0$\textit{.}

\textit{Proof:}

\textbf{Sufficiency:} We first prove that the existence of an A-type GDDW or
a B-type GDDW implies the localization of the corresponding chiral mode.

Assume we have an A-type GDDW at $x_0$. According to Definition 1, for any $%
x_L$ satisfying $a < x_L < x_0$, we have $\gamma_{global}(x_L) < 0$, and for
any $x_R$ satisfying $x_0 < x_R < b$, we have $\gamma_{global}(x_R) > 0$.
The squared modulus of the wave function is
\begin{equation}
|\psi_{k}(x)|^2=\frac{1}{\mathcal{N}^2} e^{\frac{2}{v}\int_{0}^{x}
\gamma_{global}(x^{\prime}) dx^{\prime}}>0.  \label{wave_func_suff}
\end{equation}
We now find the derivative of the squared modulus for $x$. Based on Equation (\ref{wave_func_suff}), due to $\gamma_{global}(x)$ is a
continuous function, we obtain:
\begin{equation}  \label{wave_func_der_suff}
\frac{\mathrm{d}|\psi_{k}(x)|^2}{\mathrm{d}x} = \frac{2}{ v} |\psi_{k}(x)|^2
\gamma_{global}(x).
\end{equation}
Since $\gamma_{global}(x_L) < 0$ and $\gamma_{global}(x_R) > 0$, according
to Equation (\ref{wave_func_der_suff}), we deduce that:
\begin{equation}
\begin{aligned} sign(v) \frac{\mathrm{d}|\psi_{k}(x_L)|^2}{\mathrm{d}x} &<
0, \quad \textrm{in the interval } [a, x_0), \\ sign(v)
\frac{\mathrm{d}|\psi_{k}(x_R)|^2}{\mathrm{d}x} &> 0, \quad \textrm{in the
interval } (x_0, b]. \end{aligned}
\end{equation}
This means that for the chiral mode with a negative speed, the squared modulus is monotonically increasing in the interval $[a, x_0)$ and monotonically
decreasing in the interval $(x_0, b]$. Thus, the chiral mode with negative
speed will be localized at $x_0$.

Similarly, assume we have a B-type GDDW at $x_0$. According to Definition 2,
for any $x_L$ satisfying $a < x_L < x_0$, we have $\gamma_{global}(x_L) > 0$%
, and for any $x_R$ satisfying $x_0 < x_R < b$, we have $%
\gamma_{global}(x_R) < 0$. Using the same arguments as before with the
derivative of the squared modulus in Equation (\ref{wave_func_der_suff}), we
can conclude that the squared modulus is monotonically increasing in the
interval $[a,x_0)$ and monotonically decreasing in the interval $(x_0, b]$.
Consequently, the chiral mode with positive speed will be localized at $x_0$
provided we have a B-type GDDW.

\textbf{Necessity:} Now, we prove that the localization of chiral modes can
only occur if there is an A-type or B-type GDDW.

Assume a chiral mode with negative speed is localized at $x_0$. Based on
Definition 3, the squared modulus of the wave function should be
monotonically increasing in the interval $[a,x_0]$ and monotonically
decreasing in the interval $[x_0,b]$ when $a < x_0 < b$. Thus, we know that
for $x$ in the interval $[a, x_0)$:
\begin{equation}  \label{necessity_positive}
\frac{\mathrm{d}|\psi_{k}(x)|^2}{\mathrm{d}x} > 0.
\end{equation}
Equation (\ref{necessity_positive}) implies $sign(v) \gamma_{global}(x) > 0$
for all $x$ in the interval $[a, x_0)$. Similarly, for $x$ in the interval $%
(x_0, b]$:
\begin{equation}  \label{necessity_positive_2}
\frac{\mathrm{d}|\psi_{k}(x)|^2}{\mathrm{d}x} < 0.
\end{equation}
This constraint implies $sign(v) \gamma_{global}(x) < 0$ for all $x$ in the
interval $(x_0, b]$. Combining these results, we conclude that an A-type
GDDW must exist if the chiral mode with negative speed is localized.

Likewise, if a chiral mode with negative speed is localized at $x_0$, by
following the same logic, we can conclude that a B-type GDDW must exist for
a localized chiral mode with positive speed.

\section{Non-Hermitian chiral skin effect with a pair of uniform GDDW}
\label{appendix:c}
we consider a pair of GDDW that separates the regions with
different strengths of dissipation in the left or right region. The uniform GDDW is set as
\begin{equation}
\gamma_{L}(0<x<L_{1})=\gamma_{1};\gamma_{R}(L_{1}<x<L)=\gamma_{2}
\end{equation}
where $\gamma_{R}$ and $\gamma_{L}$ are the dissipation strengths in the
right and left regions, respectively. Then we have

\begin{equation}
\bar{\gamma}=\frac{1}{L}\int_{0}^{L}\gamma(x)dx=\frac{\gamma_{1}\cdot
L_{1}+\gamma_{2}\cdot(L-L_{1})}{L}
\end{equation}

Corroding to Eq.\:(\ref{wave_func_sp}), in the left region $x\in\lbrack0,L_{1}]$%
, we have

\begin{align}
\psi^{(1)}(x) & =\frac{1}{\mathcal{N}_{1}}e^{ikx}e^{\frac{1}{v}%
\int_{0}^{x}[\gamma(x)-\bar{\gamma}]dx}  \notag \\
& =\frac{1}{\mathcal{N}_{1}}e^{ikx}e^{\alpha_{1}x}
\end{align}
where $\alpha_{1}=\frac{(\gamma_{1}-\gamma_{2})\cdot(L-L_{1})}{vL}$. In the
right region $x\in\lbrack L_{1},L]$, we have

\begin{align}
\psi^{(2)}(x) & =\frac{1}{\mathcal{N}_{2}}e^{ikx}e^{\frac{1}{v}%
\int_{L_{1}}^{x}[\gamma(x)-\bar{\gamma}]dx}  \notag \\
& =\frac{1}{\mathcal{N}_{2}}e^{ikx}e^{\alpha_{2}(x-L_{1})}
\end{align}
where $\alpha_{2}=\frac{(\gamma_{2}-\gamma_{1})\cdot L_{1}}{vL}.$Applying
the boundary conditions $\psi^{(1)}(0)=\psi^{(2)}(L),\psi^{(1)}(L_{1})=%
\psi^{(2)}(L_{1})$, we set $\mathcal{N}_{1}=e^{\alpha_{1}\cdot L_{1}},%
\mathcal{N}_{2}=1$. Finally, we have

\begin{align}
\psi^{(1)}(x) & =\frac{1}{e^{\alpha_{1}\cdot L_{1}}}e^{ikx}e^{\frac {%
(\gamma_{1}-\gamma_{2})\cdot(L-L_{1})}{vL}x} \\
\psi^{(2)}(x) & =e^{ikx}e^{\frac{(\gamma_{2}-\gamma_{1})\cdot L_{1}}{vL}%
(x-L_{1})}
\end{align}

For the case1 of L$_{1}=$L/2, $\protect\gamma_{L}(0<x<L/2)=\protect%
\gamma_{1};\protect\gamma _{R}(L/2<x<L)=\protect\gamma_{2}$, we have $\alpha_{1}=\frac{\gamma_{1}-\gamma_{2}}{2v}$ and $\alpha_{2}=%
\frac{\gamma_{2}-\gamma_{1}}{2v}$. The wave function of the edge states for
a given wavenumber $k$ in the presence of the domain wall is given by
\begin{align}
\psi^{(1)}(x) & =\frac{1}{e^{\frac{(\gamma_{1}-\gamma_{2})\cdot L}{4v}}}%
e^{ikx}e^{\frac{\gamma_{1}-\gamma_{2}}{2v}x} \\
\psi^{(2)}(x) & =e^{ikx}e^{\frac{\gamma_{2}-\gamma_{1}}{2v}(x-L/2)}
\end{align}
where $\xi$ is the localization length, which is given by%
\begin{equation}
\xi=\frac{2v}{\left\vert \gamma_{R}-\gamma_{L}\right\vert }.
\end{equation}

The following is a discussion of the specific physical implications of this example.

\textbf{Example 1:} The effective model corresponds to Fig.\:2(a) and
Fig.\:3(b) in the main text. The parameters are set as L$_{1}=$L/2, $%
\gamma_{L}(0<x<L/2)=-\gamma, \gamma _{R}(L/2<x<L)=\gamma$.

In this case, $\alpha_{1}=-\frac{\gamma}{v}$and $\alpha_{2}=\frac{\gamma}{v}$%
. The wave function of the edge states for a given wavenumber $k$ in the
presence of the domain wall is given by%
\begin{align}
\psi^{(1)}(x) & =\frac{1}{e^{\frac{\gamma L}{2v}}}e^{ikx}e^{-\frac{\gamma }{v%
}x} \\
\psi^{(2)}(x) & =e^{ikx}e^{\frac{\gamma}{v}(x-L/2)}
\end{align}
where $\xi$ is the localization length, which is given by%
\begin{equation}
\xi=\frac{v}{\gamma}.
\end{equation}

\textbf{Example 2:} The effective model corresponds to Fig.\:2(b) in the
main text. The parameters are set as L$_{1}=$L/2, $\gamma_{L}(0<x<L/2)=-%
\gamma, \gamma _{R}(L/2<x<L)=0$.\newline
In this case, $\alpha_{1}=-\frac{\gamma}{2v}$and $\alpha_{2}=\frac{\gamma}{2v%
}$. The wave function of the edge states for a given wavenumber $k$ in the
presence of the domain wall is given by
\begin{align}
\psi^{(1)}(x) & =\frac{1}{e^{\frac{\gamma L}{2v}}}e^{ikx}e^{-\frac{\gamma }{%
2v}x} \\
\psi^{(2)}(x) & =e^{ikx}e^{\frac{\gamma}{2v}(x-L/2)}
\end{align}
where $\xi$ is the localization length, which is given by%
\begin{equation}
\xi=\frac{2v}{\gamma}.
\end{equation}

\section{Non-Hermitian chiral skin effect with multiple GDDWs}
\label{appendix:d}

\begin{figure}[ptb]
\includegraphics[width=8.6cm]{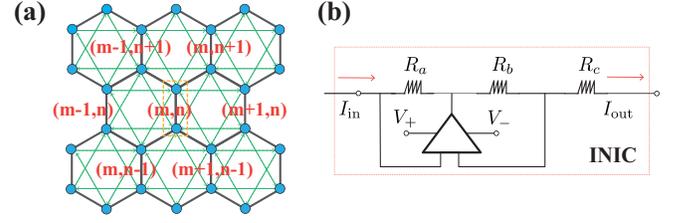}
\caption{(a)Schematic of the non-Hermitian Haldane model with uniform
dissipation in bulk in the designed circuit. (b)The schematic diagram of an electrical circuit for realizing the negative resistance.}
\label{fig5}
\end{figure}

In this part, we consider multiple uniform GDDWs that Divide into $n$
regions based on different strengths of dissipation. System total length is $%
L$, with a dissipation of $\gamma_1$ over a length of $L_1$, a dissipation
of $\gamma_2$ over a length of $L_2$, and a dissipation of $\gamma_n$ over a
length of $L_n$. The uniform GDDW is set as $\gamma(0<x<L_{1})=\gamma_{1};
\gamma(L_{1}<x<L_{2}+L_{1})=\gamma_{2}; \ldots \gamma(L-L_{n}<x<L)=\gamma_{n}
$, where $\Sigma_{n}L_n=L$. Then we have

\begin{align}
\bar{\gamma}&=\frac{1}{L}\int_{0}^{L}\gamma(x)dx \\
&=\frac{\gamma_{1}\cdot L_{1}+\gamma_{2}\cdot L_{2}+\gamma_{3}\cdot
L_{3}+\ldots+\gamma_{n}\cdot L_{n}}{L}  \notag
\end{align}
where

\begin{equation}
\alpha_{1}=-\frac{\gamma_{1}\cdot(L_1-L)+\gamma_{2}\cdot L_2+\gamma_{3}\cdot
L_3+\ldots+\gamma_{n}\cdot L_n}{vL}
\end{equation}
\begin{equation}
\alpha_{2}=-\frac{\gamma_{1}\cdot L_1+\gamma_{2}\cdot
(L_2-L)+\gamma_{3}\cdot L_3+\ldots+\gamma_{n}\cdot L_n}{vL}
\end{equation}
\begin{equation}
\alpha_{n}=-\frac{\gamma_{1}\cdot L_1+\gamma_{2}\cdot L_2+\gamma_{3}\cdot
L_3+\ldots+\gamma_{n}\cdot (L_n-L)}{vL}
\end{equation}

According to Eq.\:(\ref{wave_func_sp}), the wave function is given
\begin{equation}
\psi^{(1)}(x) =\frac{1}{\mathcal{N}_{1}}e^{ikx}e^{\alpha_1 x}
\end{equation}
\begin{equation}
\psi^{(2)}(x) =\frac{1}{\mathcal{N}_{2}}e^{ikx}e^{\alpha_2 (x-L_1)}
\end{equation}
\begin{equation}
\psi^{(n)}(x) =\frac{1}{\mathcal{N}_{n}}e^{ikx}e^{\alpha_n (x-L_{n-1})}
\end{equation}

According to continuous condition, $\psi^{(1)}(0)=\psi^{(n)}(L)$, $%
\psi^{(1)}(L_{1})=\psi^{(2)}(L_{1})$, $\psi^{(2)}(L_{2})=\psi^{(3)}(L_{2})$,
$\ldots$ ,$\psi^{(n-1)}(L_{n-1}) =\psi^{(n)}(L_{n-1})$. Finally, we can get
\begin{equation}
\mathcal{N}_{n}=e^{-\Sigma_{i=n-1,i\in\mathbb{N}^+}[\alpha_i%
\cdot(L_i-L_{i-1})]}
\end{equation}
where $\mathcal{N}_{1}=1$, and $L_0=0$.

The following is a discussion of the specific multi-GDDWs in Fig.\:\ref{fig.app_zigzag_armchair}(d).
The parameters are set as $L_{i}=L/4$ ($i \in \{1,2,3,4\}$), $\gamma_{1}(0<x<L/4)=\gamma_{\mathrm{eff}}^{\mathrm{zigzag}}=-\gamma$, $\gamma_{2}(L/4<x<L/2)=\gamma_{\mathrm{eff}}^{\mathrm{armchair}}=0$, $\gamma_{3}(L/2<x<3L/4)=\gamma_{\mathrm{eff}}^{\mathrm{zigzag}}=\gamma$, $\gamma_{4}(3L/4<x<L)=\gamma_{\mathrm{eff}}^{\mathrm{armchair}}=0$.

In this case, $\alpha_{1}=-\frac{3\gamma}{4v}$, $\alpha_{3}=\frac{3\gamma}{4v}$, and $\alpha_{2}=\alpha_{4}=0$. The wave function of the edge states for a given wavenumber $k$ in the
presence of the domain wall is given by%
\begin{align}
\psi^{(1)}(x) & =e^{ikx}e^{-\frac{3\gamma}{4v}x} \\
\psi^{(2)}(x) & =\psi^{(4)}(x) =e^{ikx}\\
\psi^{(3)}(x) & =e^{-\frac{3\gamma L}{16v}}e^{ikx}e^{\frac{3\gamma}{4v}(x-L/2)}
\end{align}
where $\xi$ is the localization length on zigzag edges, which is given by%
\begin{equation}
\xi=\frac{4v}{3\gamma}.
\end{equation}

\section{Theoretical model of the designed electric circuit}
\label{appendix:e}

In this part, we theoretically demonstrate the correspondence between the
non-Hermitian Haldane model with uniform dissipation in bulk and our
designed electric circuit. Based on Kirchhoff's law, the relationship
between current and voltage at node $m$ is described by%
\begin{widetext}
\begin{equation}
I_{m}=\left[
{\displaystyle\sum\limits_{n}}
iwC_{mn}(V_{m}-V_{n})+%
{\displaystyle\sum_{n}}
\frac{1}{iwL_{mn}}(V_{m}-V_{n})+iwC_{g}V_{m}+\frac{V_{m}}{R_{m}}\right]  .
\end{equation}
where $I_{m}$ and $V_{m}$ are the net current and voltage of node $m$ with
angular frequency being $\omega$. $L_{mn}$ is the inductance between node $m$
and node $n$. $C_{mn}$ is the capacitance between node $m$ and node $n$. The
summation is taken over all nodes, which are connected to node $m$ through an inductor or a capacitor. $C_{g}$ is the ground capacitance at node $m$.
$R_{m}$ is the resistor or negative resistor at node $m$.

In PBC with loss in bulk, each lattice site possesses 4 nodes($X^{+}$,$X^{-}$,$Y^{+}$,$Y^{-}$). In this case, the voltage and
current at the site $i$ should be written as: $V_{i}=[V_{i,X^{+}},V_{i,X^{-}%
},V_{i,Y^{+}},V_{i,Y^{-}}]^{T}$ and $I_{i}=[I_{i,X^{+}},I_{i,X^{-}}%
,I_{i,Y^{+}},I_{i,Y^{-}}]^{T}$. Additionally, each site (grounded through
$C_{g}$) is connected with other sites through two kinds of coupling: (three) NN couplings ($C_{1}$), (six) NNN couplings($C_{2}$). Also, the NNN couplings are directional-dependent, and the coupling pattern determines the sign of the geometric phase $\varphi=\pi/2$.

In this case, the Kirchhoff equation on node $X^{+}(m,n,A)$ can be expressed as:
\begin{align}
I_{m,n,X^{+}}  &  =(i\omega C_{g}+\frac{1}{i\omega L_{g}}+\frac{1}{R_{g,A}%
})V_{m,n,X^{+}}^{A}+\frac{1}{iwL}(V_{m,n,X^{+}}^{A}-V_{m,n,X^{-}}%
^{A})\nonumber\\
&  +iwC_{1}\left[  (V_{m,n,X^{+}}^{A}-V_{m,n,X^{+}}^{B})+(V_{m,n,X^{+}}%
^{A}-V_{m,n-1,X^{+}}^{B})+(V_{m,n,X^{+}}^{A}-V_{m+1,n-1,X^{+}}^{B})\right]
\nonumber\\
&  +iwC_{2}\left[  (V_{m,n,X^{+}}^{A}-V_{m+1,n-1,Y^{-}}^{A})+(V_{m,n,X^{+}%
}^{A}-V_{m,n-1,Y^{+}}^{A})+(V_{m,n,X^{+}}^{A}-V_{m,n+1,Y^{-}}^{A})\right]  .
\end{align}

The Kirchhoff equation on node $X^{-}$ can be expressed as%

\begin{align}
I_{m,n,X^{-}}  &  =(iwC_{g}+\frac{1}{iwL_{g}}+\frac{1}{R_{g,A}})V_{m,n,X^{-}%
}^{A}+\frac{1}{iwL}(V_{m,n,X^{-}}^{A}-V_{m,n,X^{+}}^{A})\nonumber\\
&  +iwC_{1}\left[  (V_{m,n,X^{-}}^{A}-V_{m,n,X^{-}}^{B})+(V_{m,n,X^{-}}%
^{A}-V_{m,n-1,X^{-}}^{B})+(V_{m,n,X^{-}}^{A}-V_{m+1,n-1,X^{-}}^{B})\right]
\nonumber\\
&  +iwC_{2}\left[  (V_{m,n,X^{-}}^{A}-V_{m+1,n-1,Y^{+}}^{A})+(V_{m,n,X^{-}%
}^{A}-V_{m,n-1,Y^{-}}^{A})+(V_{m,n,X^{-}}^{A}-V_{m,n+1,Y^{+}}^{A})\right]  .
\label{21}%
\end{align}

We assume that there are no external sources so that the current flows out of each node is zero ($I_{i}=[I_{i,X^{+}},I_{i,X^{-}},I_{i,Y^{+}},I_{i,Y^{-}%
}]^{T}=0$). In this case, the Eq. (\ref{21}) becomes%

\begin{align}
0 =  &  (iwC_{g}+\frac{1}{iwL_{g}}+\frac{1}{R_{g,A}}+3iwC_{1}+6iwC_{2}%
)V_{m,n,X^{+}}^{A}+\frac{1}{iwL}(V_{m,n,X^{+}}^{A}-V_{m,n,X^{-}}%
^{A})\nonumber\\
&  -iwC_{1}(V_{m,n,X^{+}}^{B}+V_{m,n-1,X^{+}}^{B}+V_{m+1,n-1,X^{+}}%
^{B})\nonumber\\
&  -iwC_{2}(V_{m+1,n,Y^{+}}^{A}+V_{m-1,n,Y^{-}}^{A}+V_{m-1,n+1,Y^{+}}%
^{A}+V_{m+1,n-1,Y^{-}}^{A}+V_{m,n-1,Y^{+}}^{A}+V_{m,n+1,Y^{-}}^{A}),
\label{22}%
\end{align}
and
\begin{align}
0=  &  (iwC_{g}+\frac{1}{iwL_{g}}+\frac{1}{R_{g,A}}+3iwC_{1}+6iwC_{2}%
)V_{m,n,X^{-}}^{A}+\frac{1}{iwL}(V_{m,n,X^{-}}^{A}-V_{m,n,X^{+}}%
^{A})\nonumber\\
&  -iwC_{1}(V_{m,n,X^{-}}^{B}+V_{m,n-1,X^{-}}^{B}+V_{m+1,n-1,X^{-}}%
^{B})\nonumber\\
&  -iwC_{2}(V_{m+1,n,Y^{-}}^{A}+V_{m-1,n,Y^{+}}^{A}+V_{m-1,n+1,Y^{-}}%
^{A}+V_{m+1,n-1,Y^{+}}^{A}+V_{m,n-1,Y^{-}}^{A}+V_{m,n+1,Y^{+}}^{A}).
\label{23}%
\end{align}

 For convenience, the capacitance is set as $C_{1}=t_{1}C,C_{2}%
=t_{2}C$, where $C$ acts as a reference capacitance. And we set $M_{A(B)}%
=(C_{g}-\frac{1}{w^{2}L_{g}}+\frac{1}{iwR_{g,A(B)}})/C.$ Each pair of the LC
circuit has the same resonance frequency $w_{0}=1/\sqrt{LC}.$

We simply Eq. (\ref{22}) and Eq. (\ref{23}) as:%

\begin{align}
V_{m,n,X^{+}}^{A}-V_{m,n,X^{-}}^{A} =  &  -\frac{w^{2}}{w_{0}^{2}} [
-(M_{A}+3t_{1}+6t_{2})V_{m,n,X^{+}}^{A}+t_{1}(V_{m,n,X^{+}}^{B}+V_{m,n-1,X^{+}%
}^{B}+V_{m+1,n-1,X^{+}}^{B})\nonumber\\
&  +t_{2}(V_{m+1,n,Y^{+}}^{A}+V_{m-1,n,Y^{-}}^{A}+V_{m-1,n+1,Y^{+}}%
^{A}+V_{m+1,n-1,Y^{-}}^{A}+V_{m,n-1,Y^{+}}^{A}+V_{m,n+1,Y^{-}}^{A})],
\label{24}%
\end{align}
and
\begin{align}
V_{m,n,X^{-}}^{A}-V_{m,n,X^{+}}^{A} =  &  -\frac{w^{2}}{w_{0}^{2}} [
-(M_{A}+3t_{1}+6t_{2})V_{m,n,X^{-}}^{A}+t_{1}(V_{m,n,X^{-}}^{B}+V_{m,n-1,X^{-}%
}^{B}+V_{m+1,n-1,X^{-}}^{B})\nonumber\\
&  +t_{2}(V_{m+1,n,Y^{-}}^{A}+V_{m-1,n,Y^{+}}^{A}+V_{m-1,n+1,Y^{-}}%
^{A}+V_{m+1,n-1,Y^{+}}^{A}+V_{m,n-1,Y^{-}}^{A}+V_{m,n+1,Y^{+}}^{A})].
\label{25}%
\end{align}

The voltages across the inductors are set $U_{X}=V_{X^{+}}-V_{X^{-}}%
,U_{Y}=V_{Y^{+}}-V_{Y^{-}}.$The difference between Eqs. (\ref{24}) and
(\ref{25}) yields%

\begin{align}
U_{m,n,X}^{A}  &  =-\frac{w^{2}}{2w_{0}^{2}}[-(M_{A}+3t_{1}+6t_{2}%
)U_{m,n,X}^{A}+t_{1}(U_{m,n,X}^{B}+U_{m,n-1,X}^{B}+U_{m+1,n-1,X}%
^{B})\nonumber\\
&  +t_{2}(U_{m+1,n,Y}^{A}-U_{m-1,n,Y}^{A}+U_{m-1,n+1,Y}^{A}-U_{m+1,n-1,Y}%
^{A}+U_{m,n-1,Y}^{A}-U_{m,n+1,Y}^{A})].
\end{align}

We can also derive the equations for inductor $Y$ at site A, as well as for site B following the same route:%

\begin{align}
U_{m,n,Y}^{A}  &  =-\frac{w^{2}}{2w_{0}^{2}}[-(M_{A}+3t_{1}+6t_{2}%
)U_{m,n,Y}^{A}+t_{1}(U_{m,n,Y}^{B}+U_{m,n-1,Y}^{B}+U_{m+1,n-1,Y}%
^{B})\nonumber\\
&  +t_{2}(-U_{m+1,n,X}^{A}+U_{m-1,n,X}^{A}-U_{m-1,n+1,X}^{A}+U_{m+1,n-1,X}%
^{A}-U_{m,n-1,X}^{A}+U_{m,n+1,X}^{A})],
\end{align}

\begin{align}
U_{m,n,X}^{B}  &  =-\frac{w^{2}}{2w_{0}^{2}}[-(M_{B}+3t_{1}+6t_{2}%
)U_{m,n,X}^{B}+t_{1}(U_{m,n,X}^{A}+U_{m,n+1,X}^{A}+U_{m-1,n+1,X}%
^{A})\nonumber\\
&  +t_{2}(-U_{m+1,n,Y}^{B}+U_{m-1,n,Y}^{B}-U_{m-1,n+1,Y}^{B}+U_{m+1,n-1,Y}%
^{B}-U_{m,n-1,Y}^{B}+U_{m,n+1,Y}^{B})],
\end{align}

\begin{align}
U_{m,n,Y}^{B}  &  =-\frac{w^{2}}{2w_{0}^{2}}[-(M_{B}+3t_{1}+6t_{2}%
)U_{m,n,Y}^{B}+t_{1}(U_{m,n,Y}^{A}+U_{m,n+1,Y}^{A}+U_{m-1,n+1,Y}%
^{A})\nonumber\\
&  +t_{2}(U_{m+1,n,X}^{B}-U_{m-1,n,X}^{B}+U_{m-1,n+1,X}^{B}-U_{m+1,n-1,X}%
^{B}+U_{m,n-1,X}^{B}-U_{m,n+1,X}^{B})].
\end{align}

Defining $U_{\uparrow,\downarrow}=U_{X}\pm iU_{Y},$we obtain%

\begin{align}
(3t_{1}+6t_{2}-\frac{2w_{0}^{2}}{w^{2}})U_{m,n,\uparrow}^{A}  &
=-M_{A}U_{m,n,\uparrow}^{A}+t_{1}(U_{m,n,\uparrow}^{B}+U_{m,n-1,\uparrow}%
^{B}+U_{m+1,n-1,\uparrow}^{B})\nonumber\\
&  +t_{2}(e^{-i\varphi}U_{m+1,n,\uparrow}^{A}+e^{i\varphi}U_{m-1,n,\uparrow
}^{A}+e^{-i\varphi}U_{m-1,n+1,\uparrow}^{A}\nonumber\\
&  +e^{i\varphi}U_{m+1,n-1,\uparrow}^{A}+e^{-i\varphi}U_{m,n-1,\uparrow}%
^{A}+e^{i\varphi}U_{m,n+1,\uparrow}^{A}),
\end{align}
and
\begin{align}
(3t_{1}+6t_{2}-\frac{2w_{0}^{2}}{w^{2}})U_{m,n,\uparrow}^{B}  &
=-M_{B}U_{m,n,\uparrow}^{B}+t_{1}(U_{m,n,\uparrow}^{A}+U_{m,n+1,\uparrow}%
^{A}+U_{m-1,n+1,\uparrow}^{A})\nonumber\\
&  +t_{2}(e^{i\varphi}U_{m+1,n,\uparrow}^{B}+e^{-i\varphi}U_{m-1,n,\uparrow
}^{B}+e^{i\varphi}U_{m-1,n+1,\uparrow}^{B}\nonumber\\
&  +e^{-i\varphi}U_{m+1,n-1,\uparrow}^{B}+e^{i\varphi}U_{m,n-1,\uparrow}%
^{B}+e^{-i\varphi}U_{m,n+1,\uparrow}^{B}).
\end{align}
\end{widetext}
where the geometric phase $\varphi=\pi/2$.

Consider NN coupling $U_{m,n}^{B}=e^{i\mathbf{k e}_{1}}U_{m,n}^{A}$, $%
U_{m+1,n-1}^{B}=e^{i\mathbf{ke}_{2}}U_{m,n}^{A}$, $U_{m,n-1}^{B}=e^{i\mathbf{%
ke}_{3}}U_{m,n}^{A}$, and NNN coupling $U_{m-1,n}^{A}=e^{i\mathbf{kv}%
_{1}}U_{m,n}^{A}$, $U_{m,n+1}^{A}=e^{i\mathbf{kv}_{2}}U_{m,n}^{A}$, $%
U_{m+1,n-1}^{A}=e^{i\mathbf{kv}_{3}}U_{m,n}^{A}$, we can get the independent
equation for $U_{\uparrow}$ as

\begin{align}
E\left[
\begin{array}{l}
U_{\mathbf{k},\uparrow}^{A} \\
U_{\mathbf{k},\uparrow}^{B}%
\end{array}
\right] =\left[
\begin{array}{ll}
p_{k}(\varphi)-M_{A} & T_{\mathbf{k}} \\
T_{\mathbf{k}}^{\ast} & p_{k}(-\varphi)-M_{B}%
\end{array}
\right] \left[
\begin{array}{l}
U_{\mathbf{k},\uparrow}^{A} \\
U_{\mathbf{k},\uparrow}^{B}%
\end{array}
\right].   \label{32}
\end{align}
where $E=3t_{1}+6t_{2}-\frac{2w_{0}^{2}}{w^{2}}$, $p_{k}(\varphi )=2t_{2}%
\left[ \cos(k\cdot v_{1}+\varphi)+\cos(k\cdot v_{2}+\varphi )+\cos(k\cdot
v_{3}+\varphi)\right] $, $T_{\mathbf{k}}=t_{1}(e^{i\mathbf{ke}_{1}}+e^{i%
\mathbf{ke}_{2}}+e^{i\mathbf{ke}_{3}})$, $M_{A(B)}=(C_{g}-\frac {1}{%
w^{2}L_{g}}-\frac{i}{wR_{g,A(B)}})/C$.

For convenience, the grounding capacitance is set as $C_{g}=C$, $L_{g}=L$.
As a result, in eigenfrequency $\omega_{0}$, the corresponding dissipation
is $i\gamma_{A(B)}=-M_{A(B)}=-\frac{i}{R_{g,A(B)}}\sqrt{\frac{L}{C}}$. It's
to be noted that $R_{g, A(B)}$ can be negative by using INIC, which
corresponded to gain in the non-Hermitian Haldane model.

Equations (\ref{model_ham}) and (\ref{32}) share the same noninteracting
Hamiltonian and nearly all physical quantities are defined based on the
Hamiltonian should be the same. The corresponding eigenvalue is $E$, and the eigenstate is $[U_{\mathbf{k},\uparrow}^{A},U_{\mathbf{k},\uparrow}^{B}]^{T}$%
, which can be regarded as the wave function of the Haldane model with
dissipation. Based on the consistency of the mathematical formula, it is
straightforward to infer that we can implement the Haldane model by using
designed electric circuits in Fig.\ref{fig5}.

To realize a non-Hermitian Haldane model with uniform dissipation in bulk, the schematic of the designed electric circuit is shown in Fig.\ref{fig5}(a). Electric circuits include circuit components such as capacitors, inductors, resistors, and negative impedance converters with current inversion(INIC). As shown in Fig.\ref{fig5}(b), INIC can cause the current to flow in the opposite direction to the voltage, so it can be regarded as a negative resistor. In the electric circuit system, the positive (normal) resistor can correspond to dissipation (loss), and a negative resistor can correspond to gain\cite{negative_resistance}.

\end{document}